\documentclass[10pt,twocolumn,aps,prb,floatfix,superscriptaddress,longbibliography]{revtex4-1}
\usepackage[utf8]{inputenc}
\usepackage[english]{babel}
\usepackage{graphicx}
\usepackage{epstopdf}
\usepackage{color}
\usepackage{xcolor,colortbl}
\usepackage{amssymb}
\usepackage{amsmath}
\usepackage{comment}
\usepackage{xr}

\begin{document}
\title{Vibronic states and their effect on the temperature and strain dependence of silicon-vacancy qubits in 4H silicon carbide}
\author{P\'eter Udvarhelyi}
\affiliation{Department of Biological Physics, E\"otv\"os University, 
P\'azm\'any P\'eter s\'et\'any 1/A, H-1117 Budapest, Hungary}
\affiliation{Wigner Research Centre for Physics, P.O. Box 49, H-1525 Budapest, Hungary}
\affiliation{Department of Atomic Physics, Budapest University of Technology and Economics, Budafoki \'ut 8., H-1111 Budapest, Hungary}
\author{Gerg\H{o} Thiering}
\affiliation{Wigner Research Centre for Physics, P.O. Box 49, H-1525 Budapest, Hungary}
\author{Naoya Morioka}
\affiliation{3rd Institute of Physics, University of Stuttgart and Institute for Quantum Science and Technology IQST, 70569, Stuttgart, Germany}
\author{Charles Babin}
\affiliation{3rd Institute of Physics, University of Stuttgart and Institute for Quantum Science and Technology IQST, 70569, Stuttgart, Germany}
\author{Florian Kaiser}
\affiliation{3rd Institute of Physics, University of Stuttgart and Institute for Quantum Science and Technology IQST, 70569, Stuttgart, Germany}
\author{Daniil Lukin}
\affiliation{E. L. Ginzton Laboratory, Stanford University, Stanford, CA, USA}
\author{Takeshi Ohshima}
\affiliation{National Institutes for Quantum and Radiological Science and Technology, Takasaki, Gunma 370- 1292, Japan}
\author{Jawad Ul-Hassan}
\affiliation{Department of Physics, Chemistry and Biology, Link\"oping University, SE-58183, Link\"oping, Sweden}
\author{Nguyen Tien Son}
\affiliation{Department of Physics, Chemistry and Biology, Link\"oping University, SE-58183, Link\"oping, Sweden}
\author{Jelena Vu\v{c}kovi\'c}
\affiliation{E. L. Ginzton Laboratory, Stanford University, Stanford, CA, USA}
\author{J\"org Wrachtrup}
\affiliation{3rd Institute of Physics, University of Stuttgart and Institute for Quantum Science and Technology IQST, 70569, Stuttgart, Germany}
\author{Adam Gali}
\affiliation{Wigner Research Centre for Physics, P.O. Box 49, H-1525 Budapest, Hungary}
\affiliation{Department of Atomic Physics, Budapest University of Technology and Economics, Budafoki \'ut 8., H-1111 Budapest, Hungary}
\date{\today}

\begin{abstract}
    Silicon-vacancy qubits in silicon carbide (SiC) are emerging tools in quantum technology applications due to their excellent optical and spin properties. In this paper, we explore the effect of temperature and strain on these properties by focusing on the two silicon-vacancy qubits, V1 and V2, in 4H SiC. We apply density functional theory beyond the Born-Oppenheimer approximation to describe the temperature dependent mixing of electronic excited states assisted by phonons. We obtain polaronic gap around 5 and 22~meV for V1 and V2 centers, respectively, that results in significant difference in the temperature dependent dephasing and zero-field splitting of the excited states, which explains recent experimental findings. We also compute how crystal deformations affect the zero-phonon-line of these emitters. Our predictions are important ingredients in any quantum applications of these qubits sensitive to these effects.
\end{abstract}

\maketitle

\section{Introduction}
Deep level paramagnetic point defects in solid state hosts can be utilized for quantum technology applications owing to their long living coherent spin state. In these applications, the quality of the optical properties is crucial. Additionally, these parameters are coupled to external perturbations such as changes in the strain and electromagnetic fields and temperature, either directly or elaborately mixed by spin-orbit and electron-phonon coupling. In this paper, we study these interactions in the negatively charged silicon-vacancy centers in 4H silicon carbide (SiC). These are emerging quantum defects with good spin coherence time at cryogenic temperature and fluorescence in the near infrared region~\cite{Baranov2011,Hain2014,Widmann2015,Nagy2018}. The centers have $\text{C}_{3\text{v}}$ symmetry and a spin quartet $^{4}A_{2}$ ground state~\cite{Mizuochi2005,Janzen2009, Kraus2013}. They show extremely robust zero-phonon-line (ZPL)~\cite{Nagy2019} from the excited $^{4}A_{2}$ state in the photoluminescence (PL) spectrum, labeled by V1 (1.438~eV) and V2 (1.352~eV) for $h$-site and $k$-site vacancy, respectively~\cite{Wagner2000, Ivady2017}. Another ZPL for $h$-site silicon-vacancy was observed at 5~meV higher in energy associated with the transition from the second excited $^{4}E$ state called V1$^\prime$~\cite{Janzen2009}. However, its counterpart for $k$-site silicon-vacancy (V2$^\prime$) has not yet been observed. The ground and excited state spin sublevels are split by the dipolar electron spin-electron spin interaction that is called zero-field splitting (ZFS). Due to Kramers degeneracy, only axial splitting with $2D$ is allowed between the $\pm \frac{1}{2}$ and $\pm \frac{3}{2}$ levels. The experimental values of the ZFS in the ground state are 5~MHz and 70~MHz for V1 and V2 centers, respectively, that are relatively small values because of the small deviation from the quasi tetrahedral symmetry of the spin density~\cite{Son2019}. In striking contrast, these values are in the region of about 1~GHz in the excited state for both centers at cryogenic temperatures~\cite{Nagy2019, Banks2019}, with a strong temperature dependence for the V2 center between the cryogenic and room temperature~\cite{Anisimov2016}.
 
In this paper, we wished to explore the origin of the measured magneto-optical parameters and their temperature dependence with \emph{ab initio} calculations and experimental data. We calculated the excited state polaronic fine structure of V1 and V2 PL centers using advanced density functional theory (DFT). We modelled the vibronic interactions between these states including a crystal field strain as a perturbation. From these results, we identify V1$^\prime$ and V2$^\prime$ as polaronic excited states, in contrast with the previously assumed electronic excited states. We predict the magnitude of the ZFS in the excited state as a function of temperature. We also provide a model for the temperature dependent PL linewidth of the centers upon resonant excitation.

\section{Computational methods}
We applied a screened hybrid density functional theory, Heyd-Scuzeria-Ernzerhof HSE06 DFT~\cite{HSE03, HSE06}, for the calculation of electronic states and structure relaxation as implemented in the plane wave based Vienna Ab initio Simulation Package (VASP)~\cite{VASP1, VASP2, VASP3, VASP4}. This functional produces accurate ionization and excitation energies of point defects in Group-IV semiconductors~\cite{Deak2010}. For the spin-orbit, vibrational and strain calculations, we used the computationally less demanding than HSE06 but still accurate Perdew-Burke-Ernzerhof (PBE) functional~\cite{PBE}. We used 420~eV plane wave cutoff and PAW formalism~\cite{Blochl1994}. Strain calculations were carried out with an increased plane wave cutoff of 600~eV. The model of the silicon-vacancy center was embedded in a 768-atom supercell, that is sufficiently large to use $\Gamma$-point sampling of the Brillouin-zone. We calculated the excited electronic structure using the $\Delta$SCF method~\cite{Gali2009}. For the calculation of ZFS parameters, we used the VASP PAW~\cite{Bodrog2014} implementation of dipolar electron spin-spin interaction as implemented by Martijn Marsman. The spin quantization axis in spin-orbit coupling calculations was specified to be parallel with the symmetry axis of the defect.  

\section{Sample preparation}
The $100~\mu\mathrm{m}$ thick $^{28}$Si$^{12}$C isotope enriched 4H SiC layer is grown by chemical vapour deposition (CVD) on a n-type (0001) 4H SiC substrate. The isotope purity is measured by secondary ion mass spectroscopy (SIMS) and inferred to be $^{28}\text{Si}\sim99.85\%$ and $^{12}\text{C}\sim99.98\%$. The thickness of the epitaxial layer is $\sim100~\mu\mathrm{m}$ and the surface was smoothened by chemical mechanical polishing (CMP). Current-voltage measurements at room temperature shows that the layer is n-type with a free carrier concentration of $\sim 6\cdot10^{13}~\mathrm{cm}^{-3}$, which is close to the concentration of shallow nitrogen donors of $\sim 3.5\cdot 10^{13}~\mathrm{cm}^{-3}$ determined from photoluminescence at low temperatures. Deep level transient spectroscopy measurements show that the dominant electron trap in the layer is related to the carbon vacancy with a concentration in the mid $10^{12}~\mathrm{cm}^{-3}$ range. Minority carrier lifetime mapping of the carrier shows a homogeneous carrier lifetime of $\sim0.6~\mu\mathrm{s}$. We expect the real value to be twice as high, as an optical method with high injection was used~\cite{KimotoSiC2014}. Thus, the density of all electron traps should be limited to the mid $10^{13}~\mathrm{cm}^{-3}$ range~\cite{DannoAPL2007}. Individually addressable silicon-vacancy centers were created through room temperature electron beam irradiation at 2~MeV with a fluence of $10^{13}~\mathrm{cm}^{-2}$. Some interstitial-related defects were removed by subsequent annealing at $300~^{\circ}\text{C}$ for 30~minutes. Note that the used 4H SiC sample was flipped to the side, i.e. by $90^\circ$ compared to the c-axis, such that the polarization of the excitation lasers was parallel to the $c$-axis $(E\parallel c)$ which allows to excite the V1 and V2 excited states with maximum efficiency.

To improve light extraction efficiency out of the high refractive index material $(n\approx 2.6)$, we fabricate a solid immersion using a focused ion beam milling machine (Helios NanoLab 650). The related surface contamination and modifications are subsequently removed by peroxymonosulfuric acid treatment for two hours.

A thin wire is placed next to the solid immersion lens to apply continuous radiofrequency waves to mix the spin ground states of silicon-vacancy centers. This suppresses optical spin pumping and allows for permanent observation of resonant absorption lines.

\section{Experimental setup and procedure}
All the experiments were performed at cryogenic temperatures of $4-28$~K in a Montana Instruments Cryostation. A home-built confocal microscope was used for optical excitation and subsequent fluorescence detection of single silicon vacancies. Initially, silicon-vacancy centers are identified via confocal microscopy and spectroscopy using continuous-wave off-resonant optical excitation at 730~nm.
For resonant optical excitation at 862~nm (V1 center) we use an external cavity tunable diode laser (Toptica DLC DL PRO 850). For resonant optical excitation at 916~nm (V2 center) we use Ti:Sa laser (Msquared Solstis). All measurements are referenced to a wavelength meter (High Finesse WS7-30) with about 30~MHz accuracy.

Laser light is focused onto the sample with a vacuum-compatible microscope objective (Zeiss EC Epiplan-Neofluar $100\times$, $\text{NA}=0.9$). The fluorescence emission is collected by the same microscope objective and separated from parasitic laser light by a dichroic mirror (Semrock Versa Chrome Edge). Phonon side band (PSB) fluorescence is detected using a silicon single-photon counting module (Excelitas SPCM-AQRH-W4 and AQRH-14). To obtain a resonant absorption spectrum, we first apply an off-resonant laser pulse to ensure that the vacancy center is in the desired negative charge state. Then, we perform a wavelength scan with the resonant laser and infer excitation efficiency via PSB detection. Throughout all the measurement procedure, radiofrequency waves are applied continuously.

\section{Theory}\label{sec:theory}
\begin{figure}
    \includegraphics[width=0.45\textwidth]{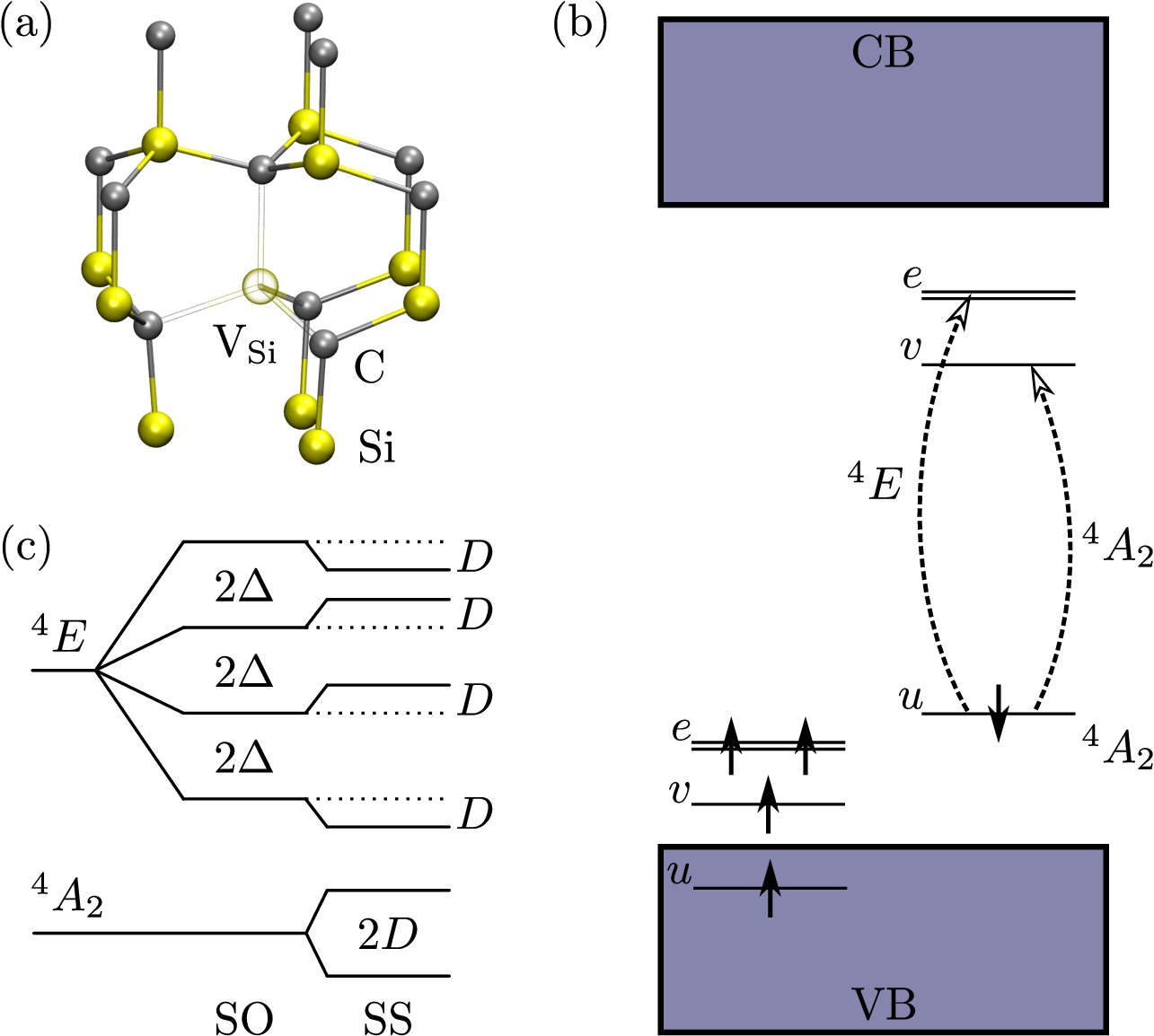}
    \caption{Geometric and electronic structure of the negatively charged silicon vacancy defect in 4H SiC. (a) Atomic model of the defect at $h$-site showing $\text{C}_{3\text{v}}$ symmetry. (b) Electronic structure of the spin-polarized $^{4}A_{2}$ ground state and the one-particle excitation schemes corresponding to $^{4}A_{2}$ and $^{4}E$ excited states. (c) Fine structure of the excited states regarding spin-orbit (SO) and dipolar spin-spin (SS) interactions (see Ref.~\onlinecite{Soykal2016}).}
    \label{fig:vacancy}
\end{figure}

Although, we are interested in the negatively charged silicon-vacancy defect in 4H SiC, it is beneficial to start the description of the structure with its counterpart in cubic (3C) polytype. In this case, the defect has $\text{T}_{\text{d}}$ point symmetry and its dangling bonds form an $a_{1}$ and a $t_{2}$ one-electron orbital. In the negatively charged state, five electrons occupy these states, two on the lower lying $a_{1}$ orbital and three on the $t_{2}$ orbital with parallel spins leading to a quartet spin ground state. The first quartet excited state can be constructed by promoting an electron from the $a_{1}$ orbital to the $t_{2}$ orbital in the spin minority channel. This excited state is Jahn-Teller unstable, its $T$ orbital symmetry will be broken by $t_{2}$ symmetric phonon modes ($\text{T}\times \text{t}_{2}$ problem)~\cite{Bersuker2006}. The hexagonal 4H polytype produces a hexagonal crystal field which statically breaks the tetrahedral symmetry. However, the key properties of the defect can be still derived from that high symmetry in the cubic crystal. This crystal field is axially symmetric in 4H SiC (called the $c$-axis), lowering the $\text{T}_{\text{d}}$ symmetry to its subgroup $\text{C}_{3\text{v}}$.  In $\text{C}_{3\text{v}}$, $t_{2}$ splits to an $a_{1}$ and an $e$ orbital. Further we call the lower lying $a_{1}$ orbital $u$ and the higher one $v$. Analogously, the ground state configuration will result in a quartet $^{4}A_{2}$. A lower excited state can be constructed by promoting an electron from the $u$ orbital to either the $v$ orbital ($^{4}A_{2}$) or the $e$ orbital ($^{4}E$). The structure of the defect and its electron configurations are depicted in Fig.~\ref{fig:vacancy}. The fine structure of the silicon-vacancy center was determined by Soykal {\it et al.}~\cite{Soykal2016}. The orbital singlets are only affected by the spin-spin interaction resulting in a splitting of the $\pm\frac{1}{2}$ and $\pm\frac{3}{2}$ Kramers doublets by $2D$ (dipolar electron spin-spin interaction). In $^{4}E$, axial spin-orbit coupling splits $8\times$ degeneracy (counting both the orbital and spin degeneracies) to $4$ Kramers doublet levels with equal energy spacing of $2\Delta$. These are also affected by the spin-spin interaction analogously to the orbital singlets.
The origin of vibronic coupling in this system can be still described by $\text{T}\times \text{t}_{2}$ Jahn-Teller problem but in the presence of the perturbing $\text{C}_{3\text{v}}$ crystal field. The potential energy surfaces of the $t_2$ orbitals are formulated using pseudo-spin of three dimension. Therefore, the vibronic interaction can be expressed on this basis as a $3 \times 3$ matrix
\begin{equation}\label{eq:W}
\mathbf{W}=\left(\begin{matrix}0 & -F_{T}Q_{\zeta} & -F_{T}Q_{\eta}\\
-F_{T}Q_{\zeta} & 0 & -F_{T}Q_{\xi}\\
-F_{T}Q_{\eta} & -F_{T}Q_{\xi} & 0
\end{matrix}\right)
    \text{,}
\end{equation}
where the orbital degrees of freedom ($t_{2}^{(\xi)}$, $t_{2}^{(\eta)}$, $t_{2}^{(\zeta)}$) are depicted by the rows and columns of the $3 \times 3$ matrix and the vibrational degrees of freedom are expressed by the $Q_{i}$ configuration coordinates. The $F_T$ linear vibronic coupling parameter connects the three $T$ vibrational normal modes with the three $t_2$ orbitals depicted by the matrix. The linear vibronic coupling can be expressed with the Jahn-Teller energy as follows (see Eq. (3.48) in Ref~[\onlinecite{Bersuker2006}])
\begin{equation}\label{F}
    F_T=\sqrt{\frac{3}{2}\hbar\omega \text{E}_\text{JT}}\text{.}
\end{equation}
$\text{C}_{3\text{v}}$ crystal field is introduced similarly to describe silicon-vacancies in 4H SiC. Thus the adiabatic potential energy surface (APES) on the $t_{2}$ basis 
\begin{equation}\label{Hamiltonian}
    \varepsilon\left(Q\right)=\frac{1}{2}\hbar\omega (Q_{\xi}^{2}+Q_{\eta}^{2}+Q_{\zeta}^{2})\mathbf{I}+\mathbf{W}-\frac{\delta}{3}\left(\begin{matrix}
    0 & 1 & 1\\
    1 & 0 & 1\\
    1 & 1 & 0 
    \end{matrix}\right)
\end{equation}
consists of the phonon energy associated to the harmonic potential of the electronic APES ($\mathbf{I}$ is the identity matrix), the vibronic interaction and the crystal field splitting ($\delta$). The associated Hamiltonian describes three coupled three-dimensional harmonic oscillators
\begin{align}
    &\hat{H}=\hbar \omega \left(\hat{a}_{\xi}^{\dagger}\hat{a}_{\xi}+\hat{a}_{\eta}^{\dagger}\hat{a}_{\eta}+\hat{a}_{\zeta}^{\dagger}\hat{a}_{\zeta}+\frac{3}{2}\right)\hat{I}\nonumber
    \\&-F\left(\hat{T}_{\xi}\hat{Q}_{\xi}+\hat{T}_{\eta}\hat{Q}_{\eta}+\hat{T}_{\zeta}\hat{Q}_{\zeta}\right)-\frac{\delta}{3}\left(\hat{T}_{\xi}+\hat{T}_{\eta}+\hat{T}_{\zeta}\right)\text{,}
\end{align}
where $\hat{a}^{\dagger}_{i}$ is the $i$ oscillator mode creation operator, $\hat{Q}_{i}=\frac{1}{\sqrt{2}}\left(\hat{a}^{\dagger}_{i}+\hat{a}_{i}\right)$ are the coordinate operators, the pseudo-spin of $t_{2}$ orbitals is represented by orbital operators
\begin{align}\label{eq:matrices}
    \hat{I}=\left(\begin{matrix}
    1 & 0 & 0\\
    0 & 1 & 0\\
    0 & 0 & 1 
    \end{matrix}\right)\text{,} && 
    \hat{T}_{\xi}=\left(\begin{matrix}
    0 & 0 & 0\\
    0 & 0 & 1\\
    0 & 1 & 0 
    \end{matrix}\right)\text{,} \nonumber\\
    \hat{T}_{\eta}=\left(\begin{matrix}
    0 & 0 & 1\\
    0 & 0 & 0\\
    1 & 0 & 0 
    \end{matrix}\right)\text{,} &&
    \hat{T}_{\zeta}=\left(\begin{matrix}
    0 & 1 & 0\\
    1 & 0 & 0\\
    0 & 0 & 0 
    \end{matrix}\right)\text{.}
\end{align}
Projections of the electronic contributions in the vibronic states to the corresponding $a_{1}$ and $e$ basis states of $\text{C}_{3\text{v}}$ symmetry are constructed using the symmetrized combinations of the $t_{2}$ orbitals.

The thermal shift and broadening of the ZPL can mainly be attributed to the vibronic interaction of quasi-degenerate orbitals, that are split by spin-orbit or crystal field interaction, with acoustic phonons. This can be modeled by time independent and time dependent perturbation theory resulting in a perturbed energy spectrum (shift) and transition rates (linewidth), respectively. 
\begin{comment}
First order perturbation in the linear vibronic coupling results linear temperature dependence in the linewidth corresponding to a single phonon resonance process~\cite{Jahnke2015}. Second order perturbation corresponds to two phonon Raman scattering processes resulting $T^{2}$ dependence in the shift and $T^{3}$ or $T^{5}$ dependence in the linewidth~\cite{Jahnke2015,Fu2009}.
\end{comment}

\section{Results and discussion}
\subsection{Polaronic spectrum in the excited state of V1 and V2 centers}
We calculate the excited state relaxed structures using constrained occupation of the Kohn-Sham levels ($\Delta$SCF method), therefore the optical excitation is straightforwardly described by promoting an electron to an unoccupied Kohn-Sham level. The calculated ZPL energies of the $^{4}A_{2}\rightarrow {^{4}A_{2}}$ optical transition for V1 and V2 centers are 1.450~eV and 1.385~eV, respectively, are in good agreement with experimental values. However, we found the second electronic excitation ($^{4}A_{2}\rightarrow {^{4}E}$) at much larger energies, 1.792~eV and 1.953~eV, respectively. The pure electronic excitation cannot account for the observed V1$^\prime$ in the PL spectrum. We concluded that the observed V1$^\prime$ state should belong to a polaronic excited state with a mixed electronic character of $^{4}A_{2}$ and $^{4}E$. In the following, we explore the polaronic spectrum and the vibronic mixing in these quantum bit defects to predict their magneto-optical properties as a function of temperature and strain.

We calculated the parameters in Eq.~\eqref{Hamiltonian} with HSE06 DFT methods using a model of silicon-vacancy center in quasi-$\text{T}_{\text{d}}$ symmetry. As the Jahn-Teller effect originates from the occupational instability of degenerate orbitals, we can relax the system to the high symmetry configuration by "smearing" the electronic occupation of the $a_{1}$ and $e$ Kohn-Sham levels corresponding to the $t_{2}$ level. The smearing eliminates the origin of symmetry breaking, thus orbital and geometric relaxation leads to the high symmetry configuration. This is done by restricting equal occupation of $v$ and $e$ orbitals at $\frac{1}{3}$ in the spin minority channel. Full self-consistent solution and structural relaxation of this constraint occupation of orbitals results in the spin quartet excited state with quasi-$\text{T}_{\text{d}}$ geometry and wavefunctions. The remaining Kohn-Sham level splitting of $v$ and $e$ orbitals in this quasi-$\text{T}_{\text{d}}$ configuration is attributed to the $\delta$ crystal field splitting. Jahn-Teller instability of this excited state results in a symmetry breaking of quasi-$\text{T}_{\text{d}}$ which leads to the two $\text{C}_{3\text{v}}$ branches that correspond to the $^{4}A_{2}$ and $^{4}E$ excited state adiabatic potential energy surfaces (APES). 

The Jahn-Teller parameters for V1 and V2 centers were calculated from the APES corresponding to quasi-$\text{T}_{\text{d}}\rightarrow \text{C}_{3\text{v}}$ distortion (see Table~\ref{tab:JTpar}). The calculated total energy difference of the high symmetry quasi-$\text{T}_{\text{d}}$ and the low symmetry $\text{C}_{3\text{v}}$ configurations equals to $E_\text{JT}$ parameter. From the high and low symmetry geometries, we calculated the normal coordinate distance of the relaxation in dimensionless units. From this result, we evaluated $\omega$ using the harmonic approximation of the APES. $F_{T}$ was obtained from Eq.~\eqref{F}.   
\begin{table}
    \caption{Jahn-Teller parameters of V1 and V2 centers obtained from DFT calculation.}
    \begin{ruledtabular}
    \begin{tabular}{cccc}
    center  &   $\text{E}_\text{JT}~(\mathrm{meV})$ &   $\hbar\omega~(\mathrm{meV})$    &   $\delta~(\mathrm{meV})$\\\hline
    V1  &   255.4   &   102.5   &   7 \\
    V2  &   396.5   &   127.5   &   29
    \end{tabular}
    \end{ruledtabular}
    \label{tab:JTpar}
\end{table}
\begin{table}
\caption{Polaronic spectrum of V1 and V2 centers and the contribution of the electronic states belonging to the corresponding irreducible representation of $\text{C}_{3\text{v}}$ point group. We define the $|\mathrm{A_{1}}\rangle\langle\mathrm{A_{1}}|$, $|\mathrm{E_{x}}\rangle\langle\mathrm{E_{x}}|$, $|\mathrm{E_{y}}\rangle\langle\mathrm{A_{y}}|$ projectors as defined by Eq.~\eqref{tab:projectordef} for the  A$_1$, E$_\text{x}$, E$_\text{y}$ contributions.  }
\begin{ruledtabular}
\begin{tabular}{cccc}
		\multicolumn{4}{c}{V1 center}\\
        rel. energy (meV)  &     A$_1$ contr. &      E$_\text{x}$ contr.   &    E$_\text{y}$ contr. \\\hline
         0.000       &     0.84      &      0.08        &    0.08      \\
         4.829       &     0.12      &      0.67        &    0.21      \\
         4.829       &     0.12      &      0.21        &    0.67      \\
        20.081       &     0.26      &      0.37        &    0.37      \\  
        66.607       &     0.53      &      0.12        &    0.35      \\
        66.607       &     0.53      &      0.35        &    0.12       
\end{tabular}
\vspace{0.5em}
\begin{tabular}{cccc}
        \multicolumn{4}{c}{V2 center} \\
        rel. energy (meV)  &     A$_1$ contr. &      E$_\text{x}$ contr.   &    E$_\text{y}$ contr. \\\hline
         0.000       &     0.94      &      0.03        &    0.03      \\
        22.070       &     0.14      &      0.21        &    0.65      \\
        22.070       &     0.14      &      0.65        &    0.21      \\
        32.824       &     0.18      &      0.41        &    0.41      \\  
        89.576       &     0.69      &      0.08        &    0.23      \\
        89.576       &     0.69      &      0.23        &    0.08   
\end{tabular}
\end{ruledtabular}\label{tab:polaronic}
\end{table}

We calculated the polaronic spectrum with the following wavefunction Ansatz:
\begin{equation}\label{tab:polaronicexpansion}
\bigl|\widetilde{\Psi}\bigr\rangle=\!\!\!\sum_{n,m,k}\!\biggl(\!c_{nmk}^{(\xi)}|t_{2}^{(\xi)}\rangle+c_{nmk}^{(\eta)}|t_{2}^{(\eta)}\rangle+c_{nmk}^{(\zeta)}|t_{2}^{(\zeta)}\rangle\!\!\biggr)|n,m,k\rangle
\text{,}
\end{equation}
where we limit our phonon or vibrational expansion up to 8th order: ($n+m+k\leq8$). Thus, in other words, we limit the $a_{\xi,\eta,\zeta}^{+}$ operators acting on the usual harmonic oscillator basis: $a_{n}^{+}|n,m,k\rangle=\sqrt{n+1}|n+1,m,k\rangle$, $a_{m}|n,m,k\rangle=\sqrt{m-1}|n,m-1,k\rangle$.
However, the symmetry of the polaronic system is reduced from $\mathrm{T_d}$ thus we project out the $A_{1}$ and $E$ electronic characters of the polaronic states to $\text{C}_{3\text{v}}$ point symmetry (see Table~\ref{tab:polaronic}) that is realized as transforming the coordinate system of Eqs.~(\ref{eq:W}),(\ref{Hamiltonian}),(\ref{eq:matrices}) towards the [111] direction of cubic $\mathrm{T_d}$ symmetry. We note that the [111] direction corresponds the $c$ axis of 4H SiC and the two $a$ and $b$ depicted by the plane perpendicular to it spanned by vectors: [$\overline{1}\overline{1}2$] and [1$\overline{1}$0] as depicted by the equations below:
\begin{equation} \label{tab:projectordef}
\begin{array}{ccccccccccc}
|\mathrm{A_{1}}\rangle & = & \!\!\bigl(\!\! &  & |t_{2}^{(\xi)}\rangle & + & |t_{2}^{(\eta)}\rangle & + & |t_{2}^{(\zeta)}\rangle & \!\!\bigr)\!\! & /\sqrt{3}\\
|\mathrm{E_{x}}\rangle & = & \!\!\bigl(\!\! & - & |t_{2}^{(\xi)}\rangle & - & |t_{2}^{(\eta)}\rangle & + & 2|t_{2}^{(\zeta)}\rangle & \!\!\bigr)\!\! & /\sqrt{6}\\
|\mathrm{E_{y}}\rangle & = & \!\!\bigl(\!\! &  & |t_{2}^{(\xi)}\rangle & - & |t_{2}^{(\eta)}\rangle &  &  & \!\!\bigr)\!\! & /\sqrt{2}
\end{array}
\text{.}
\end{equation}

Based on these results, we attribute the V1$^\prime$ ZPL line to the transition that connects the first polaronic excited state of predominantly $E$ electronic character to the electronic ground state. This way, the measured energy difference of V1 and V1$^\prime$ agrees well with the first polaronic excitation energy $\Delta_{p}\sim 5$~meV. We can similarly predict its V2$^\prime$ counterpart to be $\Delta_{p}=22$~meV higher in energy than V2. The lack of the second sharp emission in V2 center in the experiments may be explained by the relatively large energy spacing between V2 and V2$^\prime$ states because V2$^\prime$ state could be occupied at around room temperature in the PL measurements that overlaps with the sideband of the acoustic phonon modes.

From the obtained polaronic mixing of electronic states, we determined the orientation of the optical polarization associated with V1 and V1$^\prime$ transition. Using the optical selection rules, optical transition between pure electronic states $A_{2}\leftrightarrow A_{2}$ and $A_{2}\leftrightarrow E$ is allowed by $p_\parallel$ and $p_{\perp}$ polarization, respectively. $\parallel$ coincides with the $c$ axis of the crystal. As V1 and V1' transitions are the polaronic mixture of the above, the observed polarization shows an inclination from $\parallel$ direction with the angle of
\begin{equation}\label{angle}
\varphi=\arctan \left(\frac{c^{2}_{A}\mu^{2}_{A}}{c_{E}^{2}\mu_{E}^{2}}\right)\text{,} 
\end{equation}
where $c^{2}_{A}$ and $c^{2}_{E}$ are the respective $A_{1}$ and $E_{x}+E_{y}$ contributions in Table~\ref{tab:polaronic} for the corresponding polaronic state. $\mu^{2}_{A}/\mu^{2}_{E}=0.57$ relative transition dipole strength is obtained with the same method as discussed in Ref.~\onlinecite{Udvarhelyi2019}. The resulted angles of optical polarizations from Eq.~\eqref{angle} are $\varphi=18^\circ$ and $\varphi=86^\circ$ for V1 and V1' transitions, respectively. These are comparable to the experimental findings of $\sim 30^\circ$ and $\sim 90^\circ$ in Ref.~\onlinecite{Bracher2017}.

\subsection{Zero-field splitting in the polaronic excited state}\label{sec:ZFS}
\begin{figure}
    \includegraphics[width=0.45\textwidth]{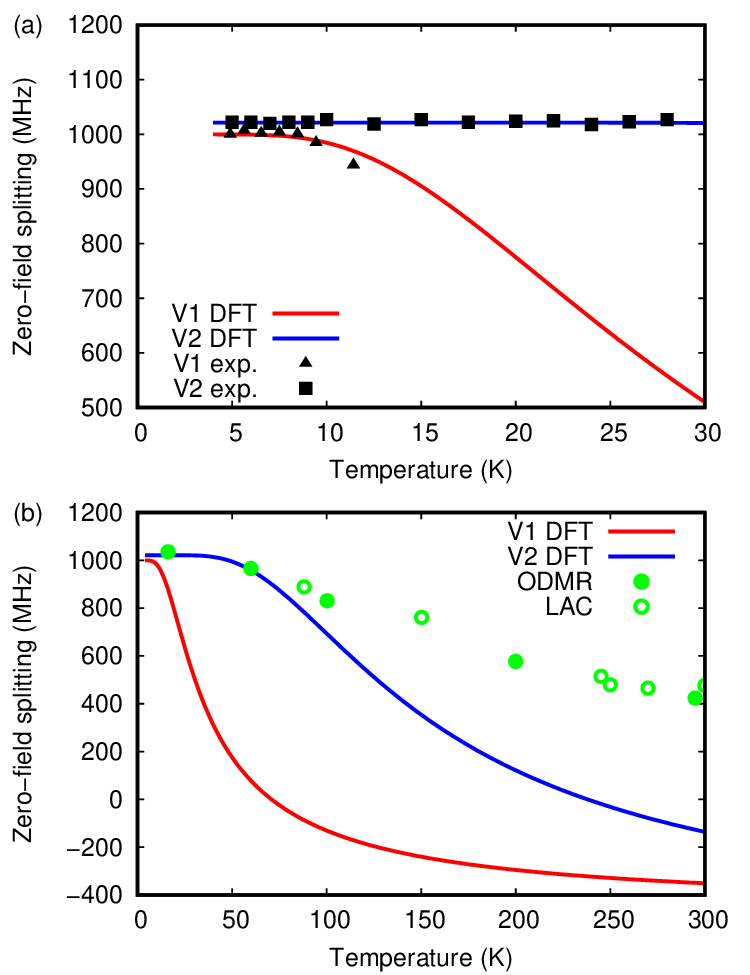}
    \caption{Temperature dependence of the excited state zero-field splitting in V1 and V2 centers (a) at cryogenic temperatures and (b) in wider temperature range. Solid curves are calculated from the Boltzmann statistics of the polaronic states with specific zero-field splitting values from the vibronic mixing. Our calculated ZFS curves are shifted by $211~\mathrm{MHz}$ (V1 DFT) and $378~\mathrm{MHz}$ (V2 DFT) for the sake of comparison to the tendencies of our observed PLE data in (a) and previous experimental data in (b) from Ref.~\onlinecite{Anisimov2016} derived from the optically detected magnetic resonance (ODMR: solid circle) and level anticrossing (LAC: hollow circle) measurements.}
    \label{fig:temperature}
\end{figure}

We use HSE06 ZFS parameters of the $^{4}A_{2}$ and $^{4}E$ electronic configurations to calculate the ZFS in polaronic states as a contribution weighted average of the electronic characters. With the calculated ZFS parameters of $2D_{^{4}A_{2}}=1664~\mathrm{MHz}$ and $2D_{^{4}E}=-1216~\mathrm{MHz}$, the first polaronic state has $2D=1211~\mathrm{MHz}$ ZFS for the V1 center. Similarly, we obtain the polaronic excited state ZFS of $2D=1400~\mathrm{MHz}$ from $2D_{^{4}A_{2}}=1569~\mathrm{MHz}$ and $2D_{^{4}E}=-1172~\mathrm{MHz}$ parameters for the V2 center. Here we neglected any spin-orbit interaction in the $^{4}E$ excited state. The validity of this approximation is detailed later in Section~\ref{sec:SOC}. We calculated the temperature dependence of ZFS for V1 and V2 centers using the Boltzmann-factor weighted average ZFS from the polaronic states at the sampling temperature values. We predict a rapid decrease for the V1 center that goes to negative values for high temperature. V2 center shows a slower transient and decrease, however, Anisimov {\it et al.}~\cite{Anisimov2016} measured an even more moderate slope that looks almost linear (see Fig. \ref{fig:temperature}). We measured the ZFS at cryogenic temperature range using PLE. The trends here follow the DFT calculations closely. Note that the calculated values are shifted in order to directly compare to the experimental data (shifting values in the caption of Fig. \ref{fig:temperature}). Our DFT calculated $D$-parameters generally overestimate compared to the experimental ones as a consequence of spin contamination. The polaronic solution carries this systematic error as well. Furthermore, our vibronic solution relies on an effective single phonon model that cannot incorporate the features of phonon sideband with acoustic phonon modes into account. This can also result in a steeper convergence to a low ZFS in the high temperature limit. At higher temperatures, the more pronounced mixing of electronic characters should lead to an averaging to $\text{T}_{\text{d}}$ symmetry where the ZFS should vanish. The experimental low and high temperature limits suggest that the intrinsic ZFS should be around $1200~\mathrm{MHz}$ and $-450~\mathrm{MHz}$ in the $^{4}A_{2}$ and $^{4}E$ excited electronic state, respectively.

\subsection{Temperature dependent PLE linewidth}
\begin{figure}
    \includegraphics[width=0.45\textwidth]{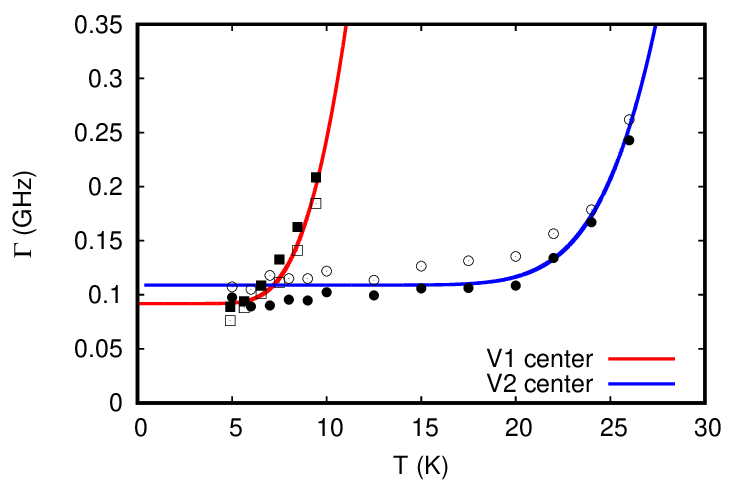}
    \caption{Temperature dependent linewidth of our observed PLE signal fit by single phonon absorption model (see Eq.~\eqref{eq:linewidth}) to the averaged $m_{S}=\pm1/2$ (hollow points) and $m_{S}=\pm3/2$ (solid points) transitions of V1 and V2 centers.}
    \label{fig:FWHM}
\end{figure}

Besides the temperature dependence of the fine structure, our polaronic solution can model the temperature dependence of the PLE linewidth as well. We expect a broadening due to the mixing of polaronic states mediated by phonons. At cryogenic temperatures, acoustic phonons dominate due to their small energy, resulting in a significant occupation number. At higher temperatures, the rapidly increasing linewidth makes the transitions between $m_S=\pm1/2$ and $m_S=\pm3/2$ indistinguishable. As the polaronic gap is quite large compared to spin-orbit splitting, we expect that only single phonon mediated processes give contribution with resonant phonon frequency to the polaronic gap. This can be formulated similarly to a resonance process in time dependent perturbation theory, where the total transition rate would be temperature dependent owing to the Bose-Einstein statistics of the acoustic phonons~\cite{Jahnke2015}. This is a temperature activated process with the specific polaronic gaps ($\Delta_{p}$) as activation energies for V1 and V2 centers. The linewidth scales with $\Delta_{p}^{3}$ owing to the product of linear and quadratic frequency scaling of the squared vibronic interaction strength and phonon density of states, respectively. We neglect the temperature dependence of $\Delta_{p}$ that would result in a higher order correction. As $\Delta_{p}$ values are much larger than the usual spin-orbit splitting of degenerate orbitals, we expect to see only the transient start of the linear dependence at cryogenic temperatures ($T<30~\mathrm{K}$). Owing to the difference in the vibronic gaps, we expect the broadening of PLE lines to start at lower temperatures for the V1 center than for the V2 center. We performed temperature dependent PLE measurements for both centers (see~Fig.~\ref{fig:FWHM}) and fit the above model for the linewidth as
\begin{equation}
\label{eq:linewidth}
    \Gamma(T)=\frac{A\Delta_{p}^{3}}{\exp{\frac{\Delta_{p}}{k_{B}T}}-1}+\Gamma_{\text{r}}+\Gamma_{1} \text{,}
\end{equation}
where the first fitting parameter $A$ incorporates the average acoustic phonon density and their coupling strength. $\Gamma_{\text{r}}=\frac{1}{2\pi\cdot 6~\mathrm{ns}}$ is the measured PL transition rate. The second fitting parameter $\Gamma_{1}$ comprises all additional temperature-independent linewidth broadening processes, such as laser power broadening and spectral diffusion. Here we assume that spectral diffusion can be described by a temperature activated process with activation energy much higher than $\Delta_{p}$, thus it has negligible temperature dependence at low temperatures. The obtained parameters after this fitting procedure are summarized in Table~\ref{tab:fit}. 
\begin{table}
    \caption{Fit parameters in the model of Eq.~\eqref{eq:linewidth}.}
    \begin{ruledtabular}
    \begin{tabular}{ccc}
    center  &   $A~(\mathrm{GHz/meV}^{3})$    &   $\Gamma_{1}~(\mathrm{GHz})$\\\hline
    V1  & $(0.37\pm0.03)$ & $(0.065\pm0.004)$  \\
    V2  & $(0.26\pm0.02)$ & $(0.082\pm0.003)$
    \end{tabular}
    \end{ruledtabular}
    \label{tab:fit}
\end{table}

\subsection{Coherence of the optical emission}
\begin{figure}
    \includegraphics[width=0.45\textwidth]{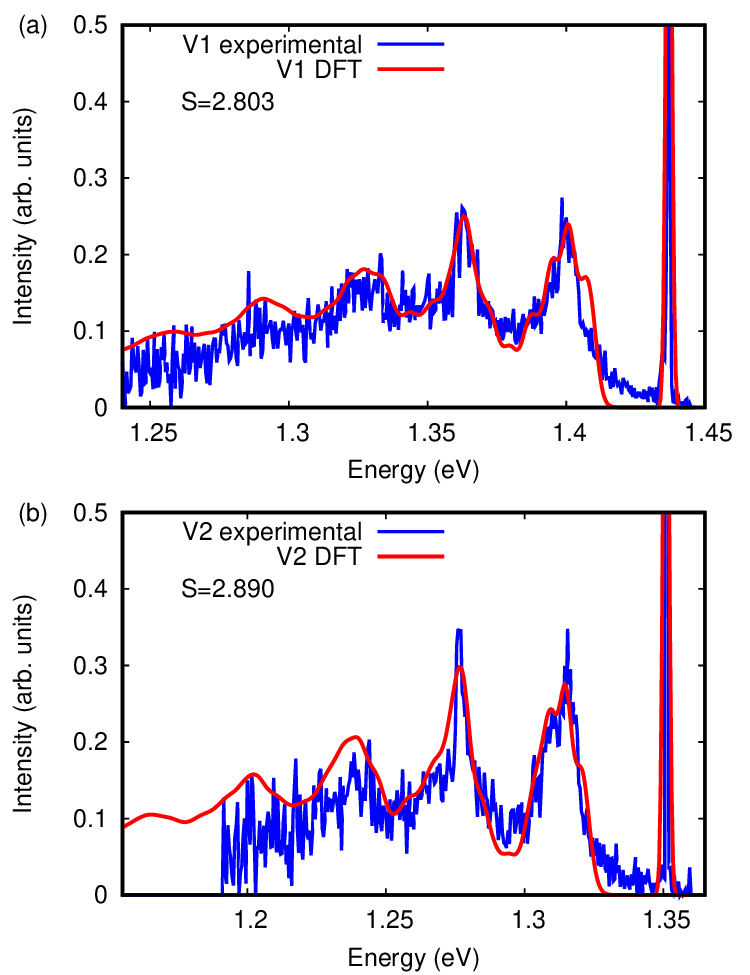}
    \caption{Experimental and calculated photoluminescence sideband of (a) V1 and (b) V2 centers. Calculated Huang-Rhys factors ($S$) are shown in the figure. All experimental data are corrected for spectrometer efficiency.}
    \label{fig:PL}
\end{figure}

After consideration the dephasing of the ZPL line caused by the acoustic phonons, we turn our attention to the overall coherence of the emission. This is characterized by the ratio of ZPL emission in the total emission, called Debye-Waller factor. In order to study this, we measured the PL signal of the silicon-vacancy centers and calculated the phonon sideband using Huang-Rhys theory~\cite{Huang1950}. The high precision $\Delta$SCF geometry optimization enables good approximation for the phonon assisted transitions. In this method, we rely on the simplified Frank-Condon principle~\cite{Alkauskas2014, Gali2016}, where we use the calculated ground state phonon spectra and normal modes. The calculation shows excellent agreement with the experimental signal (see Fig.~\ref{fig:PL}). The calculated Huang-Rhys factors ($S$) are relatively large at 0~K, and consequently the Debye-Waller factor is about 6\% for both centers. The experimental Debye-Waller factors are about 8\% and 9\% for V1 and V2 centers, respectively. Note that our measured Debye-Waller factor deviates from the previously reported value~\cite{Nagy2018}. We associate the discrepancy with the limited spectral range of the spectrometer used in the previous studies.

\subsection{The effect of spin-orbit coupling on the fine structure of the excited state}\label{sec:SOC}
\begin{figure}
    \includegraphics[width=0.45\textwidth]{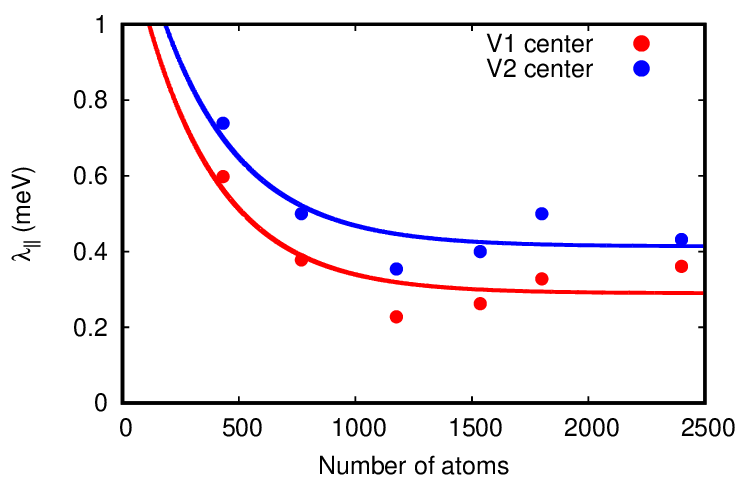}
    \caption{Size scaling of parallel spin-orbit coupling in the quasi-$\text{T}_{\text{d}}$ excited state configuration of V1 and V2 centers.}
    \label{fig:SOC}
\end{figure}

Next, we consider the effect of spin-orbit coupling in the $^{4}E$ excited state for two reasons. Firstly, we need to estimate its contribution to the zero-field splitting in the excited polaronic state. Secondly, the intrinsic spin-orbit coupling strength is an important factor in the intersystem crossing rates towards the doublet states. Therefore, we calculated the intrinsic spin-orbit coupling strength of $\lambda_{\parallel}=6\Delta=(0.29\pm 0.04)~\mathrm{meV}$ for the V1 center and $\lambda_{\parallel}=(0.41\pm 0.04)~\mathrm{meV}$ for the V2 center by calculating the spin-orbit splitting of $e$ Kohn-Sham orbitals in the quasi-$\text{T}_{\text{d}}$ excited state. These values with the corresponding standard deviation are obtained by fitting an exponential convergence for the increasing size of the supercell (see Fig.~\ref{fig:SOC}). The obtained strength of spin-orbit coupling is an order of magnitude smaller than the crystal field which implies that the perpendicular component of the spin-orbit coupling has a very weak contribution within second order perturbation theory. Furthermore, the parallel component can be considered as a small, independent perturbation to the electron-phonon system. By adding $H_\text{SO}=\lambda_{\parallel}\mathbf{LS}$  to the electron-phonon Hamiltonian in Eq.~\eqref{Hamiltonian} we obtain $20$ to $80$~MHz and $-5$ to $-20$~MHz contribution from spin-orbit coupling to the $D$-parameter for V1 center in the first ($^{4}A_{2}$) and second ($^{4}E$) polaronic branches, respectively. These values are minor corrections, as they are at least an order of magnitude smaller than the original ZFS of around 1~GHZ.  The second polaronic excited state has $\Delta=(3.3\pm1)~\mathrm{GHz}$ spin-orbit parameter which is in the expected order of magnitude~\cite{Economou2016}. In conclusion, the effect of spin-orbit coupling can be treated as a perturbation compared to the vibronic coupling, resulting in minor corrections in the polaronic spectrum and in the dipolar electron spin-spin zero-field splitting.

\subsection{Crystal strain effect on the ZPL position}
The knowledge of crystal strain effects on the optical properties of the quantum defect is also an important factor in many applications, however these are yet to be measured for the $h$ and $k$-site silicon-vacancy defects in 4H SiC. To determine these, we calculated the strain dependence of the ZPL position for normal strain components. We specified the strain matrix in the cubic reference frame where $z$ and $x$ coincides with the $[0001]$ and $[1\overline{1}00]$ directions in the hexagonal reference frame, respectively. This choice makes the $(xz)$ plane a vertical mirror plane. We deformed the 4H SiC supercell according to a single strain matrix element and allowed the defect to relax under the effect of strain. We calculate the ZPL energy for a set of 7 equidistant points in the range of $-0.003$ and $0.003$ strain, and fit the slope ($a$) to linear function. The resulting ZPL-strain coupling strengths of V1 and V2 centers are listed in Table~\ref{tab:strain}. 

\begin{figure}
    \centering
    \includegraphics[width=0.45\textwidth]{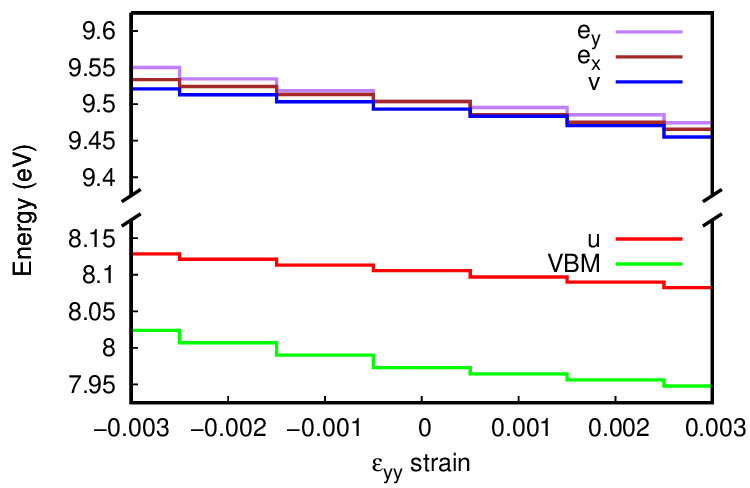}
    \caption{Strain dependence of the Kohn-Sham levels in the ground state of the V1 center. VBM is the valence band maximum, $u$, $v$ and $e$ are the defect levels depicted in Fig.~\ref{fig:vacancy} (b).}
    \label{fig:levels}
\end{figure}

These strain dependencies can be demonstrated in a simplified picture, where the changes in the ground state Kohn-Sham defect levels are followed. The analogy to the ZPL is the energy difference of the $u$ and $v$ defect level in this approximation (see Fig.~\ref{fig:vacancy} (b)). Compressive strain increases the charge density of the defect orbitals, leading to stronger repulsion between them, whereas the opposite can be stated for tensile strain. We find this closing tendency of the $u$ and $v$ energy gap in the calculations (see Fig.~\ref{fig:levels}), that can be interpreted as a negative slope in the ZPL. Furthermore, $v$ orbital is more localized on the axial carbon first neighbor of the vacancy than on the basal ones~\cite{Udvarhelyi2019}, resulting in larger response for axial strain ($a_{zz}$) compared to basal components ($a_{xx}$ and $a_{yy}$).

\begin{comment}
Relative to the valence band maximum (VBM), perpendicular strain results in a positive slope of the two energy levels, however $u$ shows larger slope than $v$. Parallel strain causes a positive slope of $u$ and negative slope of $v$. As a result, the strain dependence of the difference of the energy levels shows negative slope for all normal strain components, with a larger magnitude for $zz$ component.

The effect of strain on the individual ground and excited state total energies is reflected in the dependence of the ZPL. Strain effects the total energy of the system through the changes in the electrostatic interactions. For both compressive (negative) and dilative (positive) strain, an increase in the total energy is calculated, with compressive strain showing larger increase. For compressive (dilative) strain, the ion-ion electrostatic energy decreases (increases), whereas the sum of ion-electron and electron-electron energy increases (decreases). However, in the excited state, the former described decrease and increase are slightly smaller in magnitude, due to the slight differences in the ionic and electronic structure in the excited state. These trends results in the calculated slope of the ZPL in the function of strain.
\end{comment}

With piezoelectric actuators, a transverse tensile strain of up to $7.5\times10^{-5}$ was already demonstrated in 4H SiC~\cite{Falk2014}. This could result in around $-0.1$~meV shift in the ZPL. According to our simulations, a larger shift in ZPL can be obtained with applying a tensile strain parallel to c-axis. We note that pneumatical press on the sample can readily cause a large compressive strain in the order of $10^{-3}$ that could lead to several meVs shift of the ZPL transitions. This strong coupling to strain can be harnessed in quantum communication applications, where identical emitters are needed for efficient coupling of distant qubits with photons. An efficient control of the ZPL position compensating local differences (e.g., temperature) could be achieved by applying a specific strain using piezo actuators.

\begin{table}
    \caption{Calculated ZPL-strain coupling parameters for the V1 and V2 centers in 4H SiC.}
    \begin{ruledtabular}
    \begin{tabular}{cccc}
    center &  $a_{xx}~(\mathrm{eV/strain})$ & $a_{yy}~(\mathrm{eV/strain})$ & $a_{zz}~(\mathrm{eV/strain})$ \\\hline
    V1 & $-1.06\pm 0.08$ & $-1.41\pm 0.03$ & $-7.40\pm 0.02$ \\
    V2 & $-1.97\pm 0.09$ & $-1.89\pm 0.07$ & $-6.25\pm 0.05$ 
    \end{tabular}
    \end{ruledtabular}
    \label{tab:strain}
\end{table}

\section{Conclusion}
We investigated the negatively charged silicon-vacancy defects in 4H SiC at the hexagonal and cubic site, called V1 and V2 PL centers. We take electron-phonon interaction into account to describe their magneto-optical properties and their coupling to external perturbations. We represent this interaction based on the quasi-tetrahedral symmetry of the defect that is lowered by the crystal field of the host. We accurately reproduce the V1$^\prime$ level as a vibronic excited state and predict the V2$^\prime$ counterpart. We predict the temperature dependence of the optical linewidth and zero-field splitting that show different features for the two defects. We also determined the normal strain dependence of the ZPL position in the order of magnitude at eV/strain. Finally, we obtained the intrinsic spin-orbit coupling parameters about $0.3$ and $0.4$~meV for V1 and V2 centers, respectively, that does not contribute to the temperature dependence but are important in understanding the intersystem crossing processes. 

The physics of the vibronic coupling in the excited state is closely related to quantum applications, notably those requiring emission of highly indistinguishable photons. This comprises distribution of remote entanglement via photonic interference in a quantum repeater network~\cite{Humphreys2018}, and generation of photonic cluster states for measurement-based quantum computation~\cite{Schwartz2016, Vasconcelos2020}. The discussed interaction process with phonons in the excited state is directly related to the rate of pure dephasing of a single photon state, which affects the photon indistinguishability~\cite{Morioka2020}. Further, strain tuning of the ZPL emission, as discussed in this work, can provide an efficient means to match the photon emission wavelengths of multiple defects, or to tune emitters into resonances of optical cavities for improving photon emission rates.

\section*{Acknowledgement}
A.G.\ acknowledges the National Excellence Program of Quantum-Coherent Materials Project (Hungarian NKFIH Grant No.\ KKP129866), the EU QuantERA Q-Magine Project (Grant No.\ 127889), the QuantERA Nanospin Project (Grant No.\ 127902), the  EU H2020 Quantum Technology flagship project ASTERIQS (Grant  No.\ 820394), and the National Quantum Technology Program (Grant No.\ 2017-1.2.1-NKP-2017-00001). J.W.\ thanks the Max Planck Society as well as the EU via the ERC project SmEL. N.T.S.\ acknowledges the Swedish Research Council (Grant No.\ VR 2016-04068), J.U.H.\ thanks the Swedish Energy Agency (43611-1), N.T.S.\ and J.U.H.\ thank the Knut and Alice Wallenberg Foundation (Grant No.\ KAW 2018.0071). A.G., N.T.S.\ and J.U.H.\ thank the EU H2020 project QuanTELCO (Grant No.\ 862721). T.O.\ thanks the Japan Society for the Promotion of Science (Grant No.\ JSPS KAKENHI 17H01056 and 18H03770). J.V.\ acknowledges the US Department of Energy, Office of Science, under award No.\ DE-SC0019174. The computations were performed on resources provided by the Swedish National Infrastructure for Computing (SNIC) at NSC.

%\nocite{apsrev41Control}
%\bibliographystyle{apsrev4-1}
%\bibliography{bib}


\end{document}